\documentclass[%
 prx,
 twocolumn,
 superscriptaddress,
 amsmath,amssymb,
 longbibliography
]{revtex4-2}
\usepackage{graphicx}
\usepackage{dcolumn}
\usepackage{bm}
\usepackage{siunitx}
\usepackage{color}
\usepackage{mathtools}
\usepackage{physics}
\usepackage[colorlinks]{hyperref}
\hypersetup{%
    plainpages=true,
    breaklinks=true,
    hypertexnames=false,
    pageanchor=true,
    colorlinks=true,
    linkcolor={black},
    citecolor={blue},
    urlcolor={blue},
    anchorcolor={black}
}

\usepackage[utf8]{inputenc}
\usepackage{dsfont}
\usepackage{cleveref}
\usepackage{xcolor}
\usepackage{pgf}
\usepackage{placeins}
\usepackage{newfloat}
\newcommand{\PT}{$\mathcal{PT}$}

\begin{document}

\preprint{APS/123-QED}

\title{Probing Sensitivity near a Quantum Exceptional Point\\using Waveguide Quantum Electrodynamics}

\def\RLEaffil{Research Laboratory of Electronics, Massachusetts Institute of Technology, Cambridge, MA 02139, USA}
\def\NYUaffil{Department of Electrical and Computer Engineering, New York University, Brooklyn, NY, 11201, USA}
\def\LLaffil{MIT Lincoln Laboratory, Lexington, MA 02421, USA}
\def\Physaffil{Department of Physics, Massachusetts Institute of Technology, Cambridge, MA 02139, USA}
\def\EECSaffil{Department of Electrical Engineering and Computer Science, Massachusetts Institute of Technology, Cambridge, MA 02139, USA}
\def\affileq{These authors contributed equally.}

\author{Aziza Almanakly}
\altaffiliation{These authors contributed equally to this work.\\
\href{mailto:azizaalm@mit.edu}{azizaalm@mit.edu}
}
\affiliation{\RLEaffil}
\affiliation{\EECSaffil}

\author{R\'eouven Assouly}
\altaffiliation{These authors contributed equally to this work.\\
\href{mailto:azizaalm@mit.edu}{azizaalm@mit.edu}
}
\affiliation{\RLEaffil}

\author{Harry Hanlim Kang}
\affiliation{\RLEaffil}
\affiliation{\EECSaffil}

\author{Michael Gingras}
\author{Bethany M. Niedzielski}
\author{Hannah Stickler}
\author{Mollie E. Schwartz}
\affiliation{\LLaffil}

\author{Kyle Serniak}
\affiliation{\RLEaffil}
\affiliation{\LLaffil}

\author{Max Hays}
\affiliation{\RLEaffil}

\author{Jeffrey A. Grover}
\affiliation{\RLEaffil}
 
\author{William D. Oliver}
\email{william.oliver@mit.edu}
\affiliation{\RLEaffil}
\affiliation{\EECSaffil}
\affiliation{\Physaffil}

\begin{abstract}
 Non-Hermitian Hamiltonians with complex eigenenergies are useful tools for describing the dynamics of open quantum systems. In particular,  parity and time (\PT) symmetric Hamiltonians have generated interest due to the emergence of exceptional-point degeneracies, where both eigenenergies and eigenvectors coalesce as the energy spectrum transitions from real- to complex-valued. Because of the abrupt spectral response near exceptional points, such systems have been proposed as candidates for precision quantum sensing. In this work, we emulate a passive \PT~dimer using a two-mode, non-Hermitian system of superconducting qubits comprising one high-coherence qubit coupled to an intentionally lossy qubit via a tunable coupler. The loss is introduced by strongly coupling the qubit to a continuum of photonic modes in an open waveguide environment. Using both pulsed and continuous-wave measurements, we characterize the system dynamics near the exceptional point. We observe a behavior broadly consistent with an ideal passive \PT~dimer with some corrections due to the tunable coupler element. We extract the complex eigenenergies associated with the two modes and calculate the sensitivity as a function of the coupling strength. Confirming theoretical predictions, we observe no sensitivity enhancement near the quantum exceptional point. This work elucidates the limitations of exceptional-point systems as candidates for quantum-enhanced sensing. We establish waveguide quantum electrodynamics as a versatile platform for exploring non-Hermitian quantum dynamics in superconducting circuits.
 
\end{abstract}

\maketitle

The energies of closed quantum-mechanical systems are described by a linear operator $\mathcal H$ called the Hamiltonian. Like all other observables, the eigenvalues of the Hamiltonian must be real-valued, and its eigenvectors must span the entire Hilbert space. This naturally leads to Hermitian Hamiltonians because all Hermitian operators feature both properties. In contrast, open quantum systems can be described by non-Hermitian Hamiltonians. Certain non-Hermitian operators, such as parity and time (\PT) symmetric Hamiltonians, also feature eigenvalues that are symmetric under complex conjugation and eigenvectors that span the Hilbert space~\cite{Bender1998}.
This symmetry gives rise to two energy regimes: an unbroken regime with entirely real eigenvalues, and a regime in which \PT~symmetry is spontaneously broken with eigenvalues forming complex-conjugate pairs. The transition point between these two regimes is called the exceptional point (EP).

In both classical and quantum domains, EPs have generated interest because of their potential applications in sensing~\cite{Chen2017,Zhang2019,Shang2020}, quantum simulation~\cite{Roccati2022,Gong2022}, and topological quantum state control~\cite{Liu2021, Abbasi2022, Roccati2024}. They have been realized in a variety of platforms such as optical waveguides~\cite{Gao2008,Ruter2010,Feng2011}, optical resonators~\cite{Peng2014}, microwave cavities~\cite{Salcedo-Gallo2025}, trapped ions~\cite{Bu2023}, magnonic systems~\cite{Carney2025}, superconducting circuits~\cite{Naghiloo2019, Partanen2019, Abbasi2022, Chen2022, Teixeira2023, Erdamar2024}, and classical electronic circuits~\cite{Schindler2011,Assawaworrarit2017, Wang2020,QuirozJuarez2022}.

Because of the sharp change of system eigenenergies near exceptional points and the demonstrated utility of EPs in classical sensing, it has been suggested that EPs could offer enhanced quantum sensing capabilities~\cite{Chen2017, Ozdemir2019}. However, recent theoretical work has proven that while the sensitivity increases, the quantum noise also increases at the same rate, yielding no net signal-to-noise ratio improvement~\cite{Lau2018, Loughlin2024, Langbein2018, Ding2023, Chen2019, Schomerus2022, Darcie2025, Xu2025}. As a result, such EP systems are not expected to enhance sensitivity beyond the standard quantum limit.
The distinction between classical sensing---which can be enhanced with EPs~\cite{Chen2017}---and quantum sensing remains a point of significant tension in recent literature~\cite{Zhang2019,Shang2020}. While theoretical works have begun to converge on the inability to enhance sensitivity with quantum EPs~\cite{Lau2018, Langbein2018, Ding2023, Loughlin2024}, an unambiguous experimental demonstration in a fully quantum, tunable platform has been missing from the debate.

The canonical non-Hermitian system that features an EP is the \PT~dimer---a coupled two-mode system that exhibits parity-time symmetry. The two modes are coupled at rate $g$, with one mode exhibiting gain and the other loss, each at rate $\gamma$. The system is invariant under the joint application of the parity operator, which swaps the two modes, and the time-reversal operator, which swaps gain and loss~\cite{Ozdemir2019}. Here, we observe a generic feature of active \PT-symmetric systems: balanced loss and gain. The two eigenenergies and eigenvectors of the system coalesce at the active exceptional point where the coupling balances the loss and gain, in this case $g=\gamma / 2$.

In the case of the passive \PT-dimer variant, there is no gain but simply an imbalance between the loss of the two modes. The \PT~symmetry is broken, and the eigenvalues are always complex. However, the system can easily be transformed into the familiar \PT~dimer through a gauge transformation, which corresponds to shifting the imaginary part of the energies by the average loss rate of the two modes~\cite{Ozdemir2019}.

\begin{figure}[t!]
    \centering
    \includegraphics[width=3.45in]{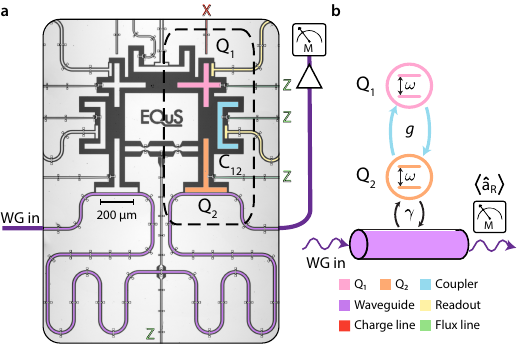}
    \caption{
    \textbf{Passive \PT~dimer experimental setup.}
    \textbf{a)} False-colored optical micrograph of the device. The system includes qubits Q$_1$ (pink) and Q$_2$ (orange); Q$_2$ is coupled to a bidirectional coplanar waveguide (purple) with strength $\gamma/2\pi = \SI{17}{MHz}$, which terminates in a measurement amplification chain. We tune the coupling $g$ between the qubits using a tunable coupler (blue) to perform measurements near the exceptional point.
    \textbf{b)} Simplified model of the system. The qubits are modeled as two-level systems that are resonant at frequency $\omega/2\pi$ = 5 GHz coupled with tunable strength $g$. We observe the physics of the system near the exceptional point by measuring the population of Q$_1$ and by probing Q$_2$ through the waveguide.}
    \label{fig:fig1}
\end{figure}




\begin{figure*}[t!]
    \centering
    \includegraphics[width=\textwidth]{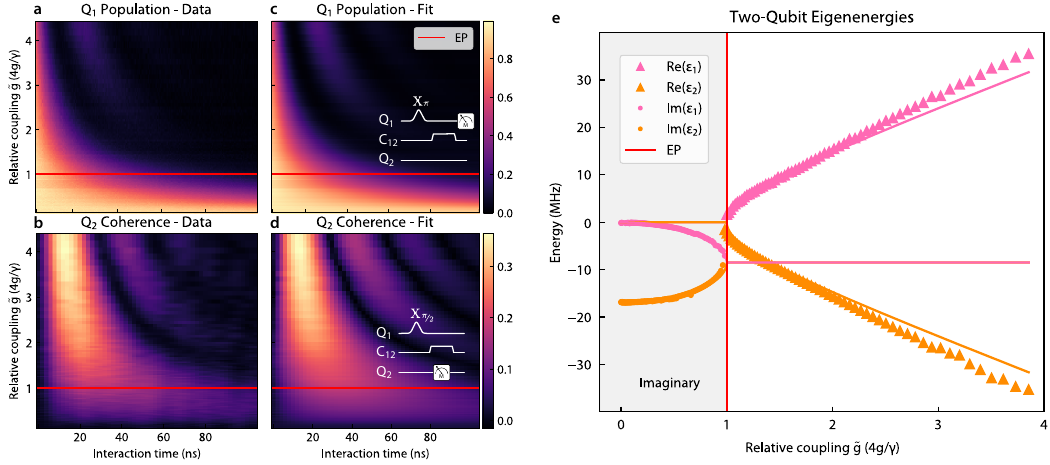}
    
    \caption{
    \textbf{Time-domain qubit measurements across the exceptional point.}
    \textbf{a)} Q$_1$ population time dynamics as a function of the relative coupling $\tilde g$. We initialize Q$_1$ in the excited state and then activate the coupling with varying strength and duration using square flux pulses applied to the tunable coupler. We measure the population using dispersive readout following the varying interaction time. Tuning the coupling strength through the EP ($g = \gamma/4$), the population time dynamics transition from exhibiting exponential decay to oscillations.
    \textbf{b)} Concurrent Q$_2$ coherence dynamics across the exceptional point. For this experiment, we initialize Q$_1$ in the $(\ket g + \ket e) / \sqrt{2}$ state, and we measure the coherence of Q$_2$ through heterodyne detection of the field-amplitude of the emission in the waveguide. We acquire the emission field signal throughout the fixed 100-ns interaction time while varying the coupling strength to traverse the EP.
    \textbf{c, d)} Each time trace is fit to theory and plotted as a function of the relative coupling for comparison. The inset shows the measurement pulse sequence.
    \textbf{e)} Measured eigenenergies of the two-qubit system as a function of the relative coupling, traversing the exceptional point. The eigenenergies are extracted from fitting the time-domain measurements. At the EP ($\tilde g = 1$, red line), the qubits' eigenenergies are degenerate. Beyond the EP, they acquire a real component while maintaining a fixed imaginary offset equal to the average loss rate $\gamma/2$. The analytical eigenenergies derived from the two-qubit non-Hermitian Hamiltonian in Eq.~\ref{eq:simpleHamiltonian} are shown with solid lines. 
    } 
    \label{fig:fig2}
\end{figure*}
In this Letter, we investigate the time dynamics and sensitivity of a passive $\mathcal{PT}$-symmetric dimer constructed from a waveguide quantum electrodynamical (wQED) system of superconducting qubits. One qubit is strongly coupled to an open waveguide, resulting in a fixed energy decay rate $\gamma$, while simultaneously coupled to a second qubit with a much smaller loss rate at a variable rate $g$ via a transmon tunable coupler~\cite{yan2018,sung2020}.
This setup enables in-situ tuning of the ratio between loss and coupling in order to bring the system from the \PT-symmetric regime to the regime of spontaneously-broken symmetry. Using the wQED architecture, we probe the complex eigenenergies while directly accessing the emission into waveguide loss channel, which is often inaccessible in EP systems~\cite{Eichler2012}.

In order to rigorously test the theoretical consensus quantum-EP sensing, we perform both time-domain pulsed and continuous-wave measurements to probe the system sensitivity to variations in the coupling strength $g$ near the EP. Consistent with theoretical predictions~\cite{Lau2018}, both measurement approaches reveal no sensitivity enhancement in the vicinity of the EP, but instead show a sensitivity peak at low coupling strength due to finite measurement time. Thus, while EPs enhance sensitivity in classical applications, our results dispel the notion that passive exceptional-point systems naturally enable quantum-enhanced sensing. By engineering dissipation in a superconducting wQED system, we demonstrate a configurable platform for studying non-Hermitian quantum dynamics.

\section{Construction of the \PT~dimer}
Our system includes two transmon qubits Q$_1$ and Q$_2$ resonant at frequency $\omega/2\pi = 5$ GHz~\cite{Koch2007}. Q$_1$ has a relatively low decay rate of \SI{29}{kHz}, while Q$_2$ has a much larger decay rate of $\gamma/2\pi = 17$ MHz due to its strong coupling to the waveguide. Because of the large disparity in loss rates, we neglect the decay of Q$_1$ in the following model. We couple the qubits via a tunable transmon coupler C$_{12}$. By changing the frequency of the tunable coupler, we control the effective coupling strength $g$ between the qubits~\cite{yan2018,sung2020}. The device micrograph and system model are shown in Figs.~\ref{fig:fig1}a and~\ref{fig:fig1}b. More details about device operation are available in Refs.~\cite{kannan2023, Almanakly2025}.

We model our system with the following non-Hermitian Hamiltonian in the rotating frame of the resonant qubits,
\begin{equation}
\frac{\hat{H}}{\hbar }=-\frac{i\gamma }{2}\hat{\sigma }_2^+\hat{\sigma }_2^-+g(\hat{\sigma }_1^+\hat{\sigma }_2^-+\hat{\sigma }_2^+\hat{\sigma }_1^-),
\label{eq:Hamiltonian}
\end{equation}
using the raising $\hat \sigma^+_1$ ($\hat \sigma^+_2$) and lowering $\hat \sigma^-_1$ ($\hat \sigma^-_2$) operators of Q$_1$ and Q$_2$. The imaginary term in the Hamiltonian represents the dissipation from Q$_2$ into the environment. Within the single-excitation subspace, the Hamiltonian simplifies to the well-known passive \PT~dimer Hamiltonian (see Supplementary Material),

\begin{equation}
    \frac{\mathcal{H}}{\hbar } =
    \begin{bmatrix}
    0 & g \\
    g & -i\gamma/2 
    \end{bmatrix},
    \label{eq:simpleHamiltonian}
\end{equation}
with complex eigenenergies $\varepsilon _{1,2}\ =\ -i\gamma/2\ \pm \ i\Gamma$, where we define $\Gamma = 2\sqrt{(\gamma/4)^2-g^2}$. This representation illustrates that this system behaves as a passive \PT~dimer due to the imbalance in dissipation between the qubits. The eigenenergies are degenerate at the passive exceptional point, where $g = \gamma/4$.

\section{\PT~dimer Eigenspectrum}
We first perform measurements of the system dynamics as a function of the relative coupling $\tilde g = 4g/\gamma$. We calibrate the frequency of the coupler such that $\tilde g = 0$ (see Supplementary Material).
We then prepare Q$_1$ into either the excited state $|\psi\rangle = |e\rangle$ or the equal superposition state $\ket\psi = (\ket g + \ket e)/\sqrt{2}$ using a resonant microwave pulse. Finally, we activate the interaction between Q$_1$ and Q$_2$ by changing the frequency of the tunable coupler using a nominally square flux pulse. To sweep the duration of the interaction, we vary the width of the square pulse in time. To control the coupling strength $\tilde g$, we vary the amplitude of the square pulse. The nonlinear mapping between amplitude and coupling $g$ is calibrated beforehand (see Supplementary Material).

Following the initialization of Q$_1$ in the excited state, we measure the population of Q$_1$ as a function of both interaction time and coupling strength using conventional dispersive readout (see Fig.~\ref{fig:fig2}a)~\cite{Wallraff2004}. We fit this time evolution to the following model, 
\begin{equation}
P_e^1(t) = e^{-\frac{\gamma}{2}t}\left| \cosh\left(\frac{\Gamma}{2}t\right) + \frac{\gamma\sinh\left(\frac{\Gamma}{2}t\right)}{2\Gamma} \right|^2,
\label{eq:data_pop}
\end{equation}
calculated from Eq.~\ref{eq:Hamiltonian} for each value of $g$. We fit each time trace of the population data to this model and plot the resulting fits in Fig.~\ref{fig:fig2}c. We plot the extracted eigenenergies $\varepsilon_1$ as a function of $\tilde g$ in Fig~\ref{fig:fig2}e.

To determine the dynamics of Q$_2$, we directly measure its microwave emission into the waveguide environment. We first initialize Q$_1$ into the equal superposition state $\ket\psi = (\ket g + \ket e)/\sqrt{2}$ using a $\pi/2$-pulse and monitor the emission of Q$_2$ into the waveguide. The waveguide terminates in an amplification chain enabling heterodyne detection~\cite{Eichler2012, kannanN00N2020}. The field amplitude maps to the Q$_2$ coherence $\expval{\hat \sigma_2^-}$ following the input-output relation~\cite{Lalumiere2013},
\begin{equation}
    \expval{\hat{a} \left(t\right)}  = \sqrt{\frac{\gamma }{2}}\expval{\hat{\sigma }_2^-(t)}. 
\end{equation}

Because we initialize Q$_1$ into the equal superposition state, which has maximal coherence, we maximize the emitted field amplitude. Figure~\ref{fig:fig2}b shows the average Q$_2$ coherence $\left| \expval{ \hat \sigma_2^-} \right|$ extracted from the emission into waveguide. The averaged qubit coherence is then fit to the following model,
\begin{equation}
\abs{\expval{\hat{\sigma }_2^-(t)}}\ =\ \frac{g}{\Gamma}e^{-\frac{\gamma }{4}t}\abs{\sinh\left(\frac{\Gamma}{2} t\right)},
\label{eq:Q2pop}
\end{equation}
in Fig.~\ref{fig:fig2}d, and the extracted eigenenergies $\varepsilon_2$ are shown in Fig.~\ref{fig:fig2}e. Although both eigenenergies may be obtained from either measurement in Figs.~\ref{fig:fig2}a and \ref{fig:fig2}b, in Fig.~\ref{fig:fig2}e we contrast the two approaches by focusing on the extraction of a single representative eigenenergy as a function of $\tilde g$.

Below the EP for $\tilde g < 1$, the eigenenergies are strictly imaginary, while beyond the EP for $\tilde g > 1$, they are complex-valued with a common imaginary shift of $-\gamma/2$, corresponding to the average loss rate of the modes. The extracted values closely follow the analytical model of the eigenenergies derived from the passive \PT~dimer Hamiltonian in Eq.~\ref{eq:simpleHamiltonian}, plotted using solid lines in Fig.~\ref{fig:fig2}e.

These equations capture the dynamics of the system close to and far from the exceptional point. For small relative coupling $\tilde g<1$, the population of Q$_1$ decays exponentially at a rate modified by the strength of the coupling $g$---the dissipation to the waveguide dominates in a phenomenon known as overdamping. For large coupling $\tilde g>1$, the qubits exchange population faster than the rate of dissipation into the waveguide, and we therefore observe oscillations in addition to the feature of exponential decay (underdamping). The exceptional point marks transition between regimes (critical damping)---on one side of the EP the system experiences only exponential decay; on the other, it exhibits an oscillatory coherent exchange between qubits that decays exponentially. 

\begin{figure}[t!]
    \centering
    \includegraphics[width=3.45in]{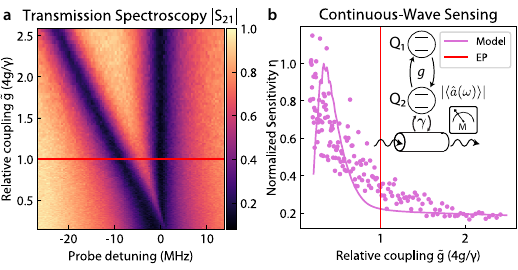}
    
    \caption{
    \textbf{Continuous-wave sensing across the exceptional point.}
    \textbf{a)} Transmission spectroscopy of the passive \PT~dimer driven by a coherent probe through the waveguide. The hybridized qubit modes split in energy according to $2g$.
    \textbf{b)} Normalized maximum sensitivity $\eta$ of the continuous-wave transmission measurement to changes in relative coupling $\tilde g$ as defined by Eq.~\ref{eq:sensitivity}. The measured observable is the emitted field amplitude $|\langle \hat a(\omega)\rangle|$.  We calculate the sensitivity for all probe detunings to determine the maximum for each $\tilde g$. We measure approximately constant noise fluctuations of the transmission signal as a function of $\tilde g$. No sensitivity improvement is observed near the EP, as predicted by three-mode, continuous-wave master equation simulations (solid purple line).
    } 
    \label{fig:fig3}
\end{figure}

\begin{figure}[t!]
    \centering
    \includegraphics[width=3.45in]{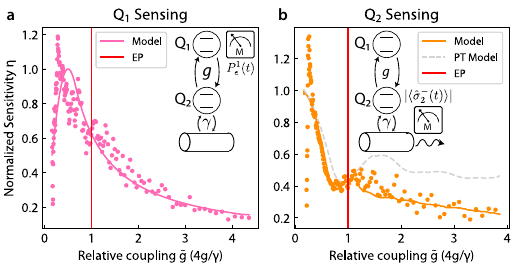}
    \caption{
    \textbf{Sensitivity across the exceptional point in the time-domain.}
    \textbf{a)} Maximum sensitivity of the Q$_1$ population measurement to variation in relative coupling $\tilde g$. The data for the measured observable $P_e^1(t)$ is shown in Fig.~\ref{fig:fig2}a. For each $\tilde g$, we calculate the sensitivity for all interaction times to determine the maximum. The noise $\sigma(t, \tilde g)$ is the standard error taken over 10,000 shots. We observe no sensitivity improvement near the EP---the decrease in sensitivity is the result of energy dissipation into the waveguide environment with increasing $\tilde g$. We model the sensitivity of the Q$_1$ population signal using the analytical expression in Eq.~\ref{eq:data_pop} (solid line).
    The decrease in sensitivity at low $\tilde g$ corresponds to the maximum sensitivity achievable within the interaction time window ($t \leq 100$ ns), as the theoretical sensitivity maximum for weak couplings occurs at times longer than the window.
    \textbf{b)} Sensitivity $\eta_{\mathrm{Q_2}}$ of the Q$_2$ coherence measurement to changes in $\tilde g$ calculated using Eq.~\ref{eq:emission_sense}.  The data for the measured coherence observable $|\langle \hat \sigma_2^-(t) \rangle|$ is shown in Fig.~\ref{fig:fig2}b. The noise over 4 million measurement shots is approximately constant as a function of $\tilde g$. We plot the sensitivity model obtained from master equation simulations of the three-mode system (solid orange line), in comparison with the analytical model of the simple \PT~dimer (dashed gray line).
    For $\tilde{g} < 0.5$, the sensitivity $\eta_\mathrm{Q_2}$ deviates from the model due to a reduced signal-to-noise ratio in the emission measurement, impacting the accuracy of the numerical derivative.
    }
    \label{fig:fig4}
\end{figure} 

\section{Sensitivity near the EP}
To evaluate the utility of the passive \PT~dimer in quantum sensing, we extract the sensitivity to small variation in the relative coupling $\tilde g$ using two different techniques: continuous-wave measurements in Fig.~\ref{fig:fig3}a and pulsed measurements in Figs.~\ref{fig:fig2}a and~\ref{fig:fig2}b, with corresponding analysis in Figs.~\ref{fig:fig4}a and~\ref{fig:fig4}b . We investigate three observables in order to sense variations in $\tilde g$: output field amplitude $\abs{\expval{\hat a (\omega)}}$, Q$_1$ population $P_e^1(t)$, and Q$_2$ coherence $\abs{\expval{\hat \sigma _2 ^- }}$. We define the unitless sensitivity $\eta(\tilde g)$ as the derivative of the observable signal with respect to the relative coupling $\tilde g$ divided by the noise fluctuation
\begin{equation}
    \eta(\tilde g) = \dv{S(\tilde g)}{\tilde g}\frac{1}{\sigma(\tilde g)},
    \label{eq:sensitivity}
\end{equation}
where $S(\tilde g)$ and $\sigma(\tilde g)$ are the mean and standard deviation of the measurement result.

In Fig.~\ref{fig:fig3}a, we measure the continuous-wave transmission spectrum $|S_{21} (\omega)| = |\langle \hat a(\omega)\rangle/\langle \hat a_\mathrm{in}(\omega)\rangle|$ as a function of probe detuning as we increase the relative coupling $\tilde g$. 
As predicted by the \PT-dimer model, the hybridized modes split by energy $2g$, while both modes adopt the average loss rate $\gamma/2$. Additionally, the tunable coupler mode imparts an approximately equal Lamb shift ($-g$) on both modes, causing the asymmetry of the dressed frequencies about zero detuning. In Fig.~\ref{fig:fig3}b, we calculate the sensitivity of the transmission signal $S(\omega, \tilde g) = S_{21}(\omega, \tilde g)$ as a function of probe frequency and relative coupling strength, highlighting the maximum sensitivity point for each coupling $\tilde g$. The noise fluctuations of this observable are approximately constant as a function of $\tilde g$. We observe a sensitivity trend consistent with three-mode master equation simulations of the continuous-wave measurement (see Supplementary Material). The frequency resolution determines the overall sensitivity maximum, which in this case occurs near $\tilde g = 1/3$. We confirm the theoretical prediction of Ref.~\cite{Lau2018}, demonstrating no sensing advantage near the EP. 


Alternatively, we approach the sensing problem using the population measurement shown in Fig~\ref{fig:fig2}a. We calculate the sensitivity of the Q$_1$ population observable signal $S(t, \tilde g) = P_e^1(t, \tilde g)$, as shown in Fig.~\ref{fig:fig4}a. The time $t$ is taken as a parameter that can be optimized for each $\tilde g$. For small coupling strengths, the measured sensitivity is limited by the finite time-resolution of our sampling interval (100 ns), which also causes the sensitivity peak observed near $\tilde g= 0.5$. This behavior is consistent with the analytical model based on Eq.~\ref{eq:data_pop}. The maximum sensitivity decreases with increasing $\tilde g$ and does not exhibit any signature of the EP. The decreasing trend of sensitivity is straightforwardly attributable to signal loss---for increasing $\tilde g$, Q$_1$ inherits more loss to the waveguide environment. Thus, we conclude that despite the sharp change in eigenenergies, we observe no sensing advantage near the EP when monitoring the Q$_1$ population. 

Similarly, we study the sensitivity of the Q$_2$ coherence observable from the measurement shown in Fig.~\ref{fig:fig2}b. We define the sensitivity of the emission measurement as
\begin{equation}
    \eta_\mathrm{Q_2} = \frac{1}{\sigma (\tilde g)} \int_0 ^{t_f} \left | \dv{|\langle \hat \sigma_2^- (t, \tilde g)\rangle|}{\tilde g} \right | dt, 
\label{eq:emission_sense}
\end{equation}
where $t_f = 100$ ns is the acquisition time. This definition reflects the fact that for each shot of the experiment, the entire time trace of the emission observable signal is acquired at once. We also compare this sensitivity to an analytical model based on the simple passive \PT~dimer in Eq.~\ref{eq:Q2pop} (dashed gray line), as well as a more complete master equation model including the tunable coupler mode (solid orange line). 
We find a much better agreement to the complete three-mode model due to the effect of the Lamb shift induced by the tunable coupler. This shift changes the frequency of the observed emission in the laboratory frame in the strong coupling regime. Finally, we observe no sensing enhancement near the EP when directly monitoring the emission to the waveguide.


\section{Conclusions}
In summary, we emulate a passive \PT~dimer using a waveguide quantum electrodynamical system of superconducting qubits. We extract the complex eigenenergies of the system while traversing the exceptional point. To probe the sensitivity of the \PT~dimer to variations in mode coupling, we perform both pulsed and continuous-wave measurements in order to study three accessible observables. Despite the abrupt spectral response near the EP, we demonstrate no sensing advantage in the near-quantum-limited sensing scenario, experimentally clarifying the theoretical debate on this subject.

Exceptional point sensing remains a useful technique in a variety of sensing scenarios limited by technical (classical) noise. For example, EP sensing schemes have been demonstrated in micromechanical silicon resonators~\cite{Zhang2024} and also proposed for optomechanical mass sensors~\cite{Djorwe2019} and various biological sensing applications~\cite{Sharma2024}. Future work will explore quantum sensing with an active \PT-symmetric dimer realized using superconducting circuits without engineered loss~\cite{WangYuXin2019,Assouly2025ExceptionalPoint}. Active \PT~dimers, where amplification is performed internally, are particularly attractive with promising demonstrations in optics~\cite{Chen2017,Liu2016}, RF~\cite{Sakhdari2022,Hajizadegan2019}, and atomic physics~\cite{Tang2023}, but are still not expected to beat the standard quantum limit. 

This passive exceptional-point sensing study demonstrates a first use case of wQED as a testbed for exploring non-Hermitian quantum dynamics. The superconducting wQED architecture enables the engineering and exploration of intricate, multi-mode open quantum systems~\cite{Ferreira2021}, including those with higher-order exceptional points~\cite{Dong2025}, transmission peak degeneracies~\cite{Carney2025}, or non-Markovian environments~\cite{Lin2025}.


\section*{Author Contributions}
AA and RA designed the experiment procedure, conducted the measurements, performed theoretical calculations and simulations, analyzed data, and wrote the manuscript.
AA designed the device.
HHK supported the data analysis.
MG, BMN, and HS fabricated the devices with coordination from KS and MES.
MH, JAG and WDO supervised the project.
All authors discussed the results and commented on the manuscript.

\section*{Acknowledgments}
The authors thank Beatriz Yankelevich for assistance with the experimental setup. 
This material is based upon work supported in part by the Air Force Office of Scientific Research Multidisciplinary University Research Initiative (MURI) under award number FA9550-22-1-0166; in part by U.S. Army Research Office Grant No. W911NF-23-1-0045; in part by the AWS Center for Quantum Computing; and in part under Air Force Contract No. FA8702-15-D-0001.
AA gratefully acknowledges support from the P.D. Soros Fellowship program and the Clare Boothe Luce Graduate Fellowship.
HHK is supported by the Korea Foundation for Advanced Studies.
Any opinions, findings, conclusions or recommendations expressed in this material are those of the author(s) and should not be interpreted as necessarily representing the official policies or endorsements of the U.S. Government.

\section*{Data Availability}
The data that support this study are available from the corresponding author upon reasonable request.

\section*{Code Availability}
The code used for numerical simulations and data analyses is available from the corresponding author upon reasonable request.

\bibliography{main}

\onecolumngrid
\newpage
\begin{center}
    \textbf{SUPPLEMENTARY INFORMATION}
\end{center}

\setcounter{figure}{0}
\setcounter{equation}{0}
\makeatletter 
\renewcommand{\thefigure}{S\@arabic\c@figure}
\renewcommand{\thetable}{S\@arabic\c@table}
\renewcommand{\theequation}{S\arabic{equation}}

\makeatother
\subsection{Device and Experimental Setup}

\newcolumntype{P}[1]{>{\centering\arraybackslash}p{#1}}

\begin{figure*}[h]
    \centering
    \includegraphics[width=\textwidth]{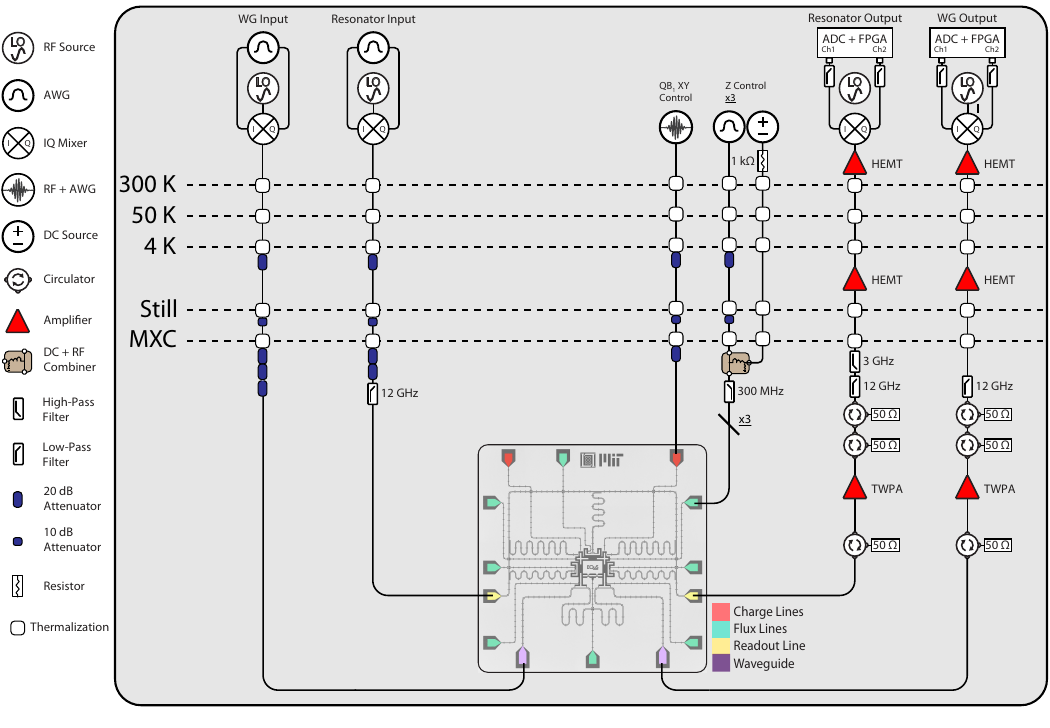}
    
    \caption{\textbf{Experimental setup.} Wiring schematic of the device and all electronics used to perform the experiment. Note that only one flux line configuration is shown (green), but each qubit and coupler is coupled to a flux line with separate, but identical, control electronics.}
    \label{fig:setup}
\end{figure*} 
This experiment was conducted in a Bluefors XLD1000 dilution refrigerator, which operates at base temperature of around 15 mK throughout the experiment.
The experimental setup is shown in Fig.~\ref{fig:setup}.
The device is protected from ambient magnetic fields by superconducting and Cryoperm-10 shields below the mixing chamber (MXC) stage.
Each end of the waveguide is connected to a microwave circulator for dual input-output operation.
To minimize thermal noise from higher temperature stages, the inputs are attenuated by 20 dB at the 4K stage, 10 dB at the Still stage, and 60 dB (40 dB for resonator readout input) at the MXC stage.
The output signals are filtered with $\SI{3}{GHz}$ high-pass and $\SI{12}{GHz}$ low-pass filters.
Two additional isolators are placed after the circulator at the MXC stage to prevent noise from higher-temperature stages travelling back into the sample. Traveling wave parametric amplifiers (TWPA) are used at the MXC stage and high electron mobility transistor (HEMT) amplifiers are used at $\SI{4}{K}$ and room-temperature stages of the measurement chain to amplify the outputs from the device.
The signals are then down-converted to an intermediate frequency using an IQ mixer, after which they are filtered, digitized, and demodulated.
Both qubits and the tunable coupler are also equipped with their own flux bias lines. A DC + RF combiner is used for all flux lines to provide both static and dynamic control of the qubit/coupler frequencies.
The DC and RF inputs are joined by a RF choke at the MXC stage before passing through a 300 MHz low-pass filter.
The RF flux control lines are attenuated by 20 dB at the $\SI{4}{K}$ stage, and by 10 dB at the 1K stage.
Q$_1$ is equipped with a local charge line for independent single-qubit XY gates.
The specific control and measurement equipment used throughout the experiment is summarized in Table~\ref{tab:equipment}.
The relevant parameters of the device used in the experiment are summarized in Table~\ref{tab:params}.

\begin{table}[h!]
\centering
\begin{tabular}{lll} 
\hline
\hline
Component & Manufacturer & Model \\
\hline
Dilution Refrigerator & Bluefors & XLD1000 \\
RF Source & Rohde \& Schwarz & SGS100A \\
DC Source & QDevil & QDAC \\
Control Chassis & Keysight & M9019A \\
AWG & Keysight & M3202A \\
ADC & Keysight & M3102A \\
\hline
\hline
\end{tabular}
\caption{\textbf{Summary of control equipment.} The manufacturers and model numbers of experimental control equipment.}
\label{tab:equipment}
\end{table}

\begin{table}[h!]
\centering
\begin{tabular}{lll} 
\hline
\hline
Parameter & $\textrm{Q}_1$ & $\textrm{Q}_2$ \\
\hline
Frequency (\si{\GHz}) & 5.0 & 5.0 \\ 
$\gamma/2\pi$ (\si{\MHz}) & - & 17 \\
$T_1$ (\si{\micro\second}) & 5.4 & - \\
$T_2^*$ (\si{\micro\second}) & 2.8 & -\\
\hline
\hline
\end{tabular}
\caption{\textbf{Summary of device parameters.} The operational qubit frequencies, qubit–waveguide coupling strength $\gamma$, and Q$_1$ coherence times $T_1$ and $T_2^*$.}
\label{tab:params}
\end{table}

\subsection{Non-Hermitian Qubit Dynamics}
To model our system as a passive \PT~dimer, we first write Eq.~\ref{eq:Hamiltonian} in matrix form in the $\{\ket{11}, \ket{01}, \ket{10}, \ket{00}\}$ basis

\begin{equation}
    \frac{\hat H_\mathrm{full}}{\hbar} = \begin{pmatrix}
        -\frac{i\gamma}{2} & 0 & 0 & 0\\
        0 & 0 & g & 0\\
        0 & g & -\frac{i\gamma}{2} & 0\\
        0 & 0 & 0 & 0
    \end{pmatrix}.
\end{equation}

\noindent Even though the system loses energy, in the non-Hermitian formulation, the $\{\ket{01}, \ket{10}\}$ manifold is stable: it is the state norm that decreases over time instead of the weight of $\ket{00}$ increasing over time.
We thus restrict ourself to that singe-photon manifold to write Eq.~\ref{eq:simpleHamiltonian}.

In the full basis, the observables corresponding to measuring the Q$_1$ excited population $M_0$ and the coherence of Q$_2$ are given by 
\begin{align*}
    M_0 &= \begin{pmatrix}
    1 & 0 & 0 & 0\\
    0 & 1 & 0 & 0\\
    0 & 0 & 0 & 0\\
    0 & 0 & 0 & 0\\
    \end{pmatrix} \\
    M_1 &= \begin{pmatrix}
    0 & 0 & 0 & 0\\
    1 & 0 & 0 & 0\\
    0 & 0 & 0 & 0\\
    0 & 0 & 1 & 0\\
    \end{pmatrix}.
\end{align*}

\noindent Using this basis, initializing Q$_1$ in $\ket e$ corresponds to the initial state $\ket{\psi_0} = (0,1,0,0)$ whereas initializing Q$_1$ in $(\ket g + \ket e)/\sqrt 2$ corresponds to $\ket{\psi_1} = (0,1/\sqrt 2,0,1/\sqrt 2)$. 
Solving for 
\[
    i\hbar \dv{\ket{\psi(t)}}{t} = \hat H_\mathrm{full} \ket{\psi(t)},
\]
we find that if the initial state is $\ket{\psi_0}$, we have
\[
    \ket{\psi(t)} = \begin{pmatrix}
        0\\
        e^{-\frac{t\gamma}{4}} \left[\cosh(\frac{\Gamma t}{2}) + \frac{\gamma}{2\Gamma} \sinh(\frac{\Gamma t}{2}) \right]\\
        2i\frac{g}{\Gamma} e^{-\frac{t\gamma}{4}} \sinh(\frac{\Gamma t}{2})\\
        0 \end{pmatrix}.
\]
The Q$_1$ excited state population is then calculated to be
\[
    P_e^1(t) = \bra{\psi(t)} M_0 \ket{\psi(t)} = e^{-\frac{\gamma}{2}t}\left| \cosh\left(\frac{\Gamma}{2}t\right) + \frac{\gamma\sinh\left(\frac{\Gamma}{2}t\right)}{2\Gamma} \right|^2.
\]
This result also matches the solution to 
\[
    i\hbar \dv{\phi_0(t)}{t} = \mathcal H \phi_0(t),
\]
with $\phi_0(t=0) = \left(1, 0\right)$ and an observable $\mathcal M_0 = \begin{pmatrix}
    1 & 0\\
    0 & 0
\end{pmatrix}$.
When the initial state is $\ket{\psi_1}$, we similarly have
\[
    \ket{\psi(t)} = \begin{pmatrix}
        0\\
        \frac{1}{\sqrt 2} e^{-\frac{t\gamma}{4}} \left[\cosh(\frac{\Gamma t}{2}) + \frac{\gamma}{2\Gamma} \sinh(\frac{\Gamma t}{2}) \right]\\
        \sqrt2 i\frac{g}{\Gamma} e^{-\frac{t\gamma}{4}} \sinh(\frac{\Gamma t}{2})\\
        1/\sqrt 2 \end{pmatrix}.
\]
The Q$_2$ coherence is calculated with
\[
    \abs{\expval{\hat{\sigma }_2^-(t)}} = \abs{\bra{\psi(t)} M_1 \ket{\psi(t)}} = \frac{g}{\Gamma}e^{-\frac{\gamma }{4}t}\abs{\sinh\left(\frac{\Gamma}{2}t\right)},
\]
which also equals the second component of the state vector $\phi_1(t)$ obtained by solving
\[
    i\hbar \dv{\phi_1(t)}{t} = \mathcal H \phi_1(t),
\]
given the initial condition $\phi_1(t=0) = \left(\frac 1 2, 0\right)$.

\newpage
\begin{figure}[t!]
    \centering
    \includegraphics[width=7in]{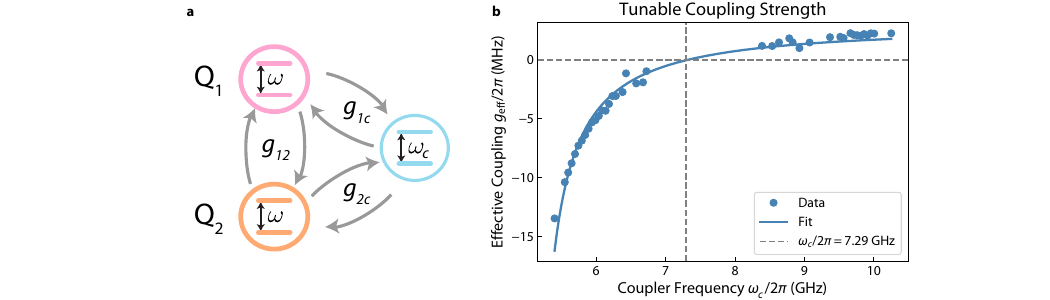}
    \caption{
    \textbf{Tunable coupling model.}
    \textbf{a)} Schematic of the three-mode system used to realize a passive \PT~dimer. The two qubits are resonant at frequency $\omega/2\pi = 5$ GHz, with a qubit-qubit coupling $g_{12}$. Each qubit is coupled to the tunable coupler at frequency $\omega_c$ at rates $g_{1c}$ and $g_{2c}$.
    \textbf{b)} Effective coupling between qubits as a function of tunable coupler frequency $g_\mathrm{eff}(\omega_c)$. We extract the effective coupling by measuring avoided level crossings while varying the coupler frequency. The extracted couplings are fit to the analytical model in Eq.~\ref{eq:tunable_g} for resonant qubits $\omega_1 = \omega_2 = 5.0$ GHz. The resulting fit parameters are reference coupler frequency $\omega_{c,\mathrm{ref}}/2\pi = 7.25$ GHz, coupling strengths $g_{12} = 5.9$ MHz, $g_\mathrm{1c,\mathrm{ref}}/2\pi = 112.4$ MHz, and $g_\mathrm{2c,\mathrm{ref}}/2\pi = 101.2$ MHz. }
    \label{fig:tunable_g}
\end{figure}
\subsection{Tunable Coupling}
The system used throughout this experiment to emulate a passive \PT~dimer comprises three modes, each modeled as a two-level system: two resonant qubit modes at frequency $\omega = \omega_1 = \omega_2$ and a tunable coupler mode at frequency $\omega_c$. The two qubits are coupled to each other at rate $g_{12}$, and each qubit is also coupled to the tunable coupler at rates $g_{1c}$ and $g_{2c}$, as depicted in Fig.~\ref{fig:tunable_g}a. The Hamiltonian for this three-mode system in the rotating frame of the resonant qubits is
\begin{equation}
    \hat H = \frac{\delta}{2}\hat \sigma_z^c + g_{12}(\hat \sigma_1^+\hat \sigma_2^- + \hat\sigma_1^-\hat \sigma_2^+ ) + g_{1c}(\hat \sigma_1^+\hat \sigma_c^- + \hat\sigma_1^-\hat \sigma_c^+ ) + g_{2c}(\hat \sigma_2^+\hat \sigma_c^- + \hat\sigma_2^-\hat \sigma_c^+ ),
    \label{eq:supp_H}
\end{equation}
where $\delta = \omega - \omega_c < 0$ is the negative qubit-coupler detuning, and $\hat \sigma_j^\pm$ are the raising and lowering operators for each qubit/coupler. We assume that the coupler remains in the ground state and all three couplings are weak compared to the qubit-coupler detuning $\delta$, enabling the application of the Schreiffer-Wolf transformation using the unitary
\begin{equation}
\hat{U} = \exp\left[ \sum_{j=1,2} \left[\frac{g_{jc}}{\delta} \left( \hat{\sigma}_j^+ \hat{\sigma}_c^- - \hat{\sigma}_j^- \hat{\sigma}_c^+ \right) - \frac{g_{jc}}{\Sigma} \left( \hat{\sigma}_j^+ \hat{\sigma}_c^+ - \hat{\sigma}_j^- \hat{\sigma}_c^- \right) \right]\right],
\end{equation}
where we define $\Sigma = \omega + \omega_c$. This transformation decouples the coupler from the system up to second order in $g_{jc}/\delta$ while accounting for counter-rotating terms~\cite{yan2018}. The resulting effective interaction Hamiltonian is
\begin{equation}
    \hat{H} = \sum_{j = 1,2} \lambda_j(\omega_c)\hat \sigma_z^j +  g_\mathrm{eff}(\omega_c) 
    \left( \hat{\sigma}_1^+ \hat{\sigma}_2^- + \hat{\sigma}_1^- \hat{\sigma}_2^+ \right),
\end{equation}
which includes the Lamb shift on the qubit frequencies,
\begin{equation}
     \lambda_j(\omega_c) = g_{jc}^2\left(\frac{1}{\delta} - \frac{1}{\Sigma}\right),
\end{equation}
and the expression for the effective tunable coupling between the qubits, 
\begin{equation}
g_\mathrm{eff}(\omega_c) = g_{1c}g_{2c}\left(\frac{1}{\delta} - \frac{1}{\Sigma}\right) + g_{12} = \, g_{1c,\mathrm{ref}} \; g_{2c,\mathrm{ref}} \; \frac{\omega_c}{\omega_{c,\mathrm{ref}}} \left( \frac{1}{\delta} - \frac{1}{\Sigma} \right) + g_{12}.
\label{eq:tunable_g}
\end{equation}
It is evident from these expressions that both the Lamb shift and the effective coupling exhibit a similar dependence on the coupler frequency. In this experiment, the qubits are resonant at \(\omega_1/2\pi = \omega_2/2\pi = 5~\mathrm{GHz}\), and we define a reference coupler frequency \(\omega_{c, \mathrm{ref}}/2\pi = 7.29~\mathrm{GHz}\)---the frequency at which the qubits idle effectively decoupled, with \(g_\mathrm{eff}(\omega_{c,\mathrm{ref}}) = 0\) MHz. The measured qubit-coupler couplings \((g_{1c,\mathrm{ref}}, g_{2c,\mathrm{ref}})\) are referenced to this idle point.

Using qubit spectroscopy, we directly determine $g_\mathrm{eff}(\omega_c)$ by measuring the avoided level crossings between the qubits as a function of coupler frequency $\omega_c$. We fit the extracted coupling curve to Eq.~\ref{eq:tunable_g} to determine $g_{12}, g_{1c},$ and $g_{2c}$. The tunable coupling curve of this system is shown in Fig.~\ref{fig:tunable_g}b.

This tunable coupling curve provides a calibrated, nonlinear map between coupler frequency, coupler flux bias, and effective qubit-qubit coupling. We use this same calibration throughout the data analysis to establish the nonlinear relationship between the voltage amplitude of square flux pulses applied to the coupler, the resulting in situ shift of the coupler frequency, and the corresponding effective coupling $g_\mathrm{eff}$. Throughout this work, we define $g = g_\mathrm{eff}$.

After biasing the coupler at the idle point where $g_\mathrm{eff} = 0$, we turn on the interaction by applying a square flux pulse with a voltage amplitude chosen to produce a desired value of $g_\mathrm{eff}$ according to the calibrated tunable coupling curve. The pulse duration sets the interaction time but does not affect the coupling calibration, as the effective coupling is treated as static throughout the square pulse.

\newpage
\subsection{Continuous-Wave Sensing Model}
To model the continuous-wave sensing approach, we employ the master-equation formalism for the three-mode system driven with a resonant probe through the waveguide
\begin{equation}
    \partial_t \hat{\rho} = -i\big[\hat{H_\mathrm{d}},\hat\rho\big] + \gamma D\big[\hat{\sigma}_2^-\big]\hat\rho,
\end{equation}
where $\gamma$ is the coupling of Q$_2$ to the waveguide, and $D[\hat{O}] = \hat{O}\hat\rho\hat{O}^\dagger - \frac{1}{2} \{\hat{O}^\dagger\hat{O},\hat\rho\}$ is the Lindblad dissipator. We write the driven Hamiltonian in the rotating frame of the probe
\begin{equation}
    \hat H_d = \frac{\delta_p}{2}(\hat \sigma_z^1 + \hat \sigma_z^2) + \frac{\omega_c - \omega_p}{2}\hat \sigma_z^c + \frac{\Omega_p}{2}\hat \sigma_x^2
    + g_{12}(\hat \sigma_1^+\hat \sigma_2^- + \hat\sigma_1^-\hat \sigma_2^+ ) + g_{1c}(\hat \sigma_1^+\hat \sigma_c^- + \hat\sigma_1^-\hat \sigma_c^+ ) + g_{2c}(\hat \sigma_2^+\hat \sigma_c^- + \hat\sigma_2^-\hat \sigma_c^+ ). 
\end{equation}
where $\delta_p = \omega - \omega_p$ is the qubit-probe detuning, and $\Omega_\mathrm{p}$ is the drive strength of the probe. Given a rightward-propagating probe input, the output from the right end of the waveguide is~\cite{Lalumiere2013}
\begin{equation}
    \hat a_\mathrm{R} = \hat a_\mathrm{R}^\textrm{in} + \sqrt{\frac{\gamma}{2}}\hat \sigma_2^{-}.
    \label{eq:supp_IO}
\end{equation}

\begin{figure}[t!]
    \centering
    \includegraphics[width=7in]{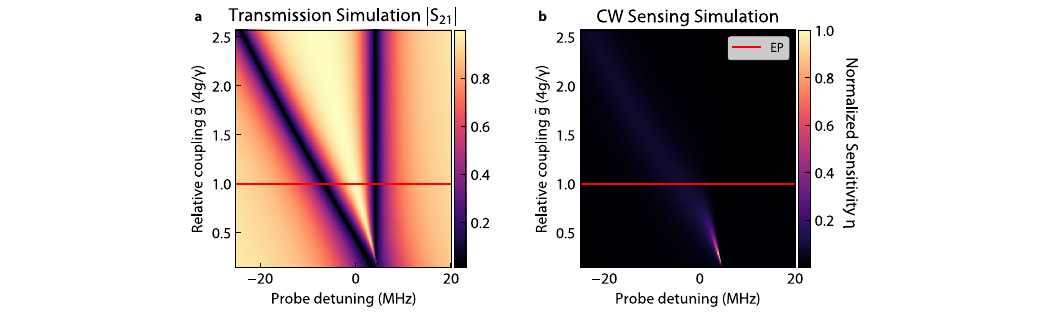}
    \caption{
    \textbf{Continuous-wave sensing model.}
    \textbf{a)} Steady-state master equation simulation of the three-mode system. We drive the system with a coherent probe through the waveguide and vary the relative coupling $\tilde g$ between the qubits. We observe an increasing splitting of the modes of the passive \PT~dimer along with the commensurate Lamb shift. The hybridized modes inherit an equal decay rate $\gamma/2$ to the waveguide.
    \textbf{b)} Simulation of the continuous-wave sensing experiment. We take the derivative of the transmission simulation with respect to $\tilde g$ for all probe detunings. The maximum normalized sensitivity for each $\tilde g$ is used as the model trace in  Fig.~\ref{fig:fig3}b.} 
    \label{fig:cw_sim}
\end{figure}

To simulate the transmission spectrum, $S_{21}(\delta_p) = \langle \hat a_\mathrm{R}(\delta_p) \rangle/ \langle \hat a_\mathrm{R}^\textrm{in}(\delta_p)\rangle$, we numerically solve the master equation in the steady state, setting $\partial_t\hat \rho = 0$, to determine $\langle \hat \sigma_2^{-}(\delta_p) \rangle$ while varying the frequency of the tunable coupler $\omega_c$, effectively changing the coupling between the qubits. We plot the results of the transmission simulation as a function of coupling in Fig.~\ref{fig:cw_sim}a.

We model the sensitivity of the transmission signal to changes in relative coupling $\tilde g$ by numerically taking the derivative of the transmission simulation with respect to $\tilde g$. We plot the normalized sensitivity for all probe detunings in Fig.~\ref{fig:cw_sim}b. The model trace plotted in Fig.~\ref{fig:fig3}b is the maximum simulated sensitivity for each $\tilde g$.   

\newpage
\begin{figure}[t!]
    \centering
    \includegraphics[width=7in]{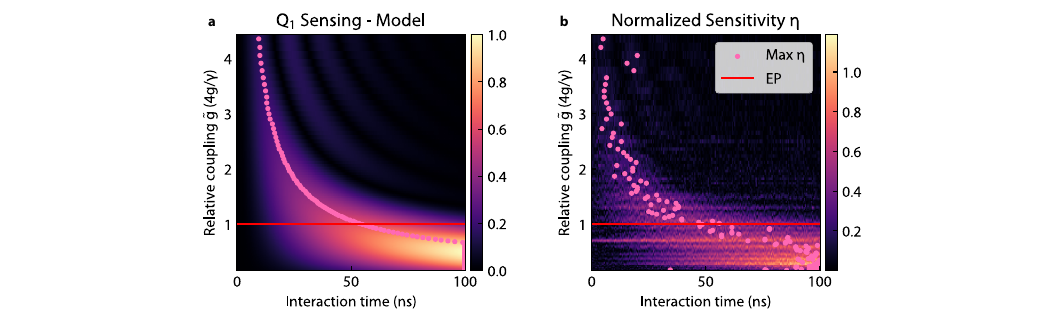}
    \caption{
    \textbf{Q$_1$ sensing model.}
    \textbf{a)} Sensing model using the analytical expression for the Q$_1$ population $P_e^1(t, \tilde g)$ as a function of time $t$ and $\tilde g$. This model is derived from the simple passive \PT~dimer Hamiltonian. The derivative along the $\tilde g$ is take for each point in the time trace, corresponding to a different square voltage pulse duration activating the interaction between qubits. The maximum sensitivity is highlighted with pink dots---these points correspond to the model trace shown in Fig.~\ref{fig:fig4}a.
    \textbf{b)} Corresponding sensitivity experiment using the measured qubit Q$_1$ population. The maximum sensitivity for each relative coupling $\tilde g$ are highlighted with pink dots---these points correspond to the data trace shown in Fig.~\ref{fig:fig4}a.} 
    \label{fig:q1_sim}
\end{figure}

\subsection{Q$_1$ Sensing Model}

We model the Q$_1$ sensing approach in Fig.~\ref{fig:fig4}a using the analytical expression for qubit population
\begin{equation}
    P_e^1(t) = e^{-\frac{\gamma}{2}t}\left| \cosh\left(\frac{\Gamma}{2}t\right) + \frac{\gamma\sinh\left(\frac{\Gamma}{2}t\right)}{2\Gamma} \right|^2,
\end{equation}
plotted in Fig.~\ref{fig:q1_sim}a to match the parameter range in the experiment shown in Fig.~\ref{fig:fig2}a.

We differentiate the analytical population time traces along the $\tilde g$ axis, as shown in Fig.~\ref{fig:q1_sim}b. We find close agreement to the sensitivity of the measured population data. Each point in the time trace corresponds to a different interaction time, or square voltage pulse width implemented in the experiment. Thus, to find the measurement settings that yield the maximum sensitivity, we calculate the sensitivity of the population for each interaction time $t$. The maximum sensitivity for each relative coupling $\tilde g$ is indicated with pink dots---this is the data plotted in Fig.~\ref{fig:fig4}a. The model curve is also extracted in the same manner using the analytical expression for the population directly---this is plotted as the solid pink line in Fig.~\ref{fig:fig4}a.

The observed maximum sensitivity for small relative couplings near $\tilde g$ is a consequence of finite sampling time. Here, the increasing decay rate of Q$_1$ is more apparent at longer interactions times outside the experimental range. For small relative couplings, the signal is maximized, as the increased decoherence rate is not yet dominating the dynamics.

Because noise is amplified by differentiation, numerical derivatives of noisy data are notoriously challenging to obtain reliably. To mitigate the effect of noise, it is common to smooth the averaged data prior to numerical differentiation. For all three sensing approaches, we apply a Gaussian smoothing filter along the $\tilde g$ axis. The effects of noise are particularly pronounced for small relative couplings $\tilde g$ where there is little change in the system dynamics. Accordingly, we choose the width of the Gaussian filter to be inversely proportional to the magnitude of $\tilde g$ for all three sensing approaches. 
\newpage

\subsection{Q$_2$ Sensing Model}
To model the sensing approach using the emission of Q$_2$ into the waveguide, we use the master equation
\begin{equation}
    \partial_t \hat{\rho} = -i\big[\hat{H},\hat\rho\big] + \gamma D\big[\hat{\sigma}_2^-\big]\hat\rho,
\end{equation}
using the Hamiltonian from Eq.~\ref{eq:supp_H}, which is written in the the rotating frame of the resonant qubits. 

To effectively tune the qubit-qubit coupling in simulation, we vary the tunable coupler frequency following the tunable coupling map presented in Fig.~\ref{fig:tunable_g}b. We calculate the average coherence of Q$_2$ as a function of relative coupling and interaction time $\langle \hat \sigma_2^-(t, \tilde g)\rangle$. Small high-frequency oscillations from the qubit–coupler detuning term in the Hamiltonian are not resolved in the measurements; therefore, we model the data by applying a Butterworth low-pass filter with an 80 MHz cutoff to the simulation results. 

In practice, we measure the emitted field amplitude $\langle\hat a_\mathrm{R}(t, \tilde g)\rangle$, which maps to the Q$_2$ coherence through the input-output relation in Eq.~\ref{eq:supp_IO}. For each relative coupling $\tilde g$ in Fig.~\ref{fig:fig2}b, the entire time trace is acquired at once for an interaction time of 100 ns. Thus, we naturally have information for all interaction times following signal acquisition. This fact justifies the redefinition of sensitivity,
\begin{equation}
    \eta_\mathrm{Q_2} = \frac{1}{\sigma (\tilde g)} \int_0 ^{t_f} \left | \dv{\langle \hat \sigma_2^- (t, \tilde g)\rangle}{\tilde g} \right | dt, 
    \label{eq:supp_q2_sense}
\end{equation}
incorporating sensing information throughout the entire acquisition window.

\begin{figure}[t!]
    \centering
    \includegraphics[width=7in]{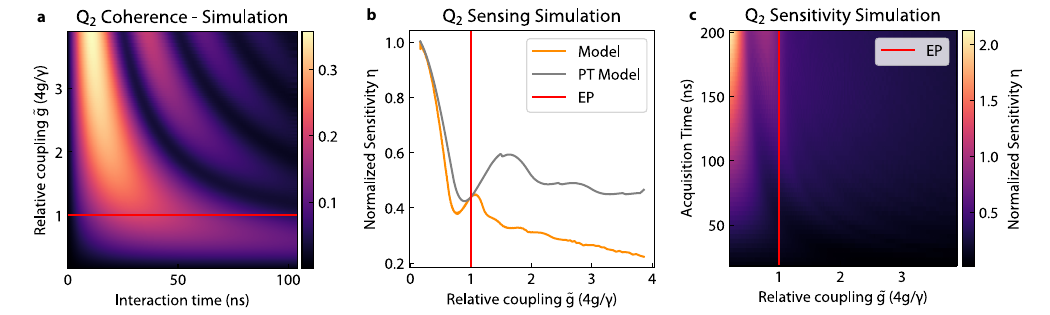}
    \caption{
    \textbf{Q$_2$ sensing model.}
    \textbf{a)} Master equation simulation using a three-mode system model. We calculate the Q$_2$ coherence as a function of coupling strength $\tilde g$ and interaction time $|\langle \hat \sigma_2^-(t,\tilde g)\rangle|$.  
    \textbf{b)} Calculated sensitivity using the three-mode model according to Eq.~\ref{eq:supp_q2_sense} (orange). We compare the sensitivity of the analytical coherence plotted in Fig.~\ref{fig:fig2}d derived from the simple \PT~dimer Hamiltonian (gray).
    \textbf{c)} Three-mode master equation simulation of the emission field sensitivity as a function of acquisition time. Throughout the experiment, the emission signal acquisition time is fixed at 100 ns. This specific time window results in a local maximum in sensitivity near the EP, as seen in the orange curve of part b. We show the location of this maximum on the relative coupling axis shifts as a function of acquisition time, confirming it as a windowing artifact rather than an inherent feature of the EP.}
    \label{fig:q2_sim}
    
\end{figure}

Using the three-mode simulation shown in Fig.~\ref{fig:q2_sim}a, we calculate the sensitivity using this redefinition, plotting the results with a solid orange line in Fig.~\ref{fig:q2_sim}b. We compare this sensitivity curve to an analytical model obtained from the simple \PT~dimer Hamiltonian. Using the analytical expression for Q$_2$ coherence,
\begin{equation}
\abs{\expval{\hat{\sigma }_2^-(t)}}\ =\ \frac{g}{\Gamma}e^{-\frac{\gamma }{4}t}\abs{\sinh\left(\frac{\Gamma}{2} t\right)},
\end{equation}
 where $\Gamma = 2\sqrt{(\frac{\gamma}{4})^2-g^2}$, we calculate and plot the sensitivity with a solid gray line in Fig.~\ref{fig:q2_sim}b. Though any differences between the time traces of the analytical fit in Fig.~\ref{fig:fig2}d and the three-mode simulation in Fig.~\ref{fig:q2_sim}a are imperceptible by eye, we find the three-mode model is clearly more consistent with the data. The three-mode model more accurately captures the effect of the tunable coupler.

\newpage

\end{document}